# Steganalysis of Image with Adaptively Parametric Activation

Hai Su, Meiyin Han, Junle Liang and Songsen Yu

*Abstract*—Steganalysis as a method to detect whether image contains secret message, is a crucial study avoiding the imperils from abusing steganography. The point of steganalysis is to detect the weak embedding signals which is hardly learned by convolutional layer and easily suppressed. In this paper, to enhance embedding signals, we study the insufficiencies of activation function, filters and loss function from the aspects of reduce embedding signal loss and enhance embedding signal capture ability. Adaptive Parametric Activation Module is designed to reserve negative embedding signal. For embedding signal capture ability enhancement, we add constraints on the high-pass filters to improve residual diversity which enables the filters extracts rich embedding signals. Besides, a loss function based on contrastive learning is applied to overcome the limitations of cross-entropy loss by maximum inter-class distance. It helps the network make a distinction between embedding signals and semantic edges. We use images from BOSSbase 1.01 and make stegos by WOW and S-UNIWARD for experiments. Compared to state-of-the-art methods, our method has a competitive performance. Source code is available via GitHub: https://github.com/MyRoe/deep_learning_detector

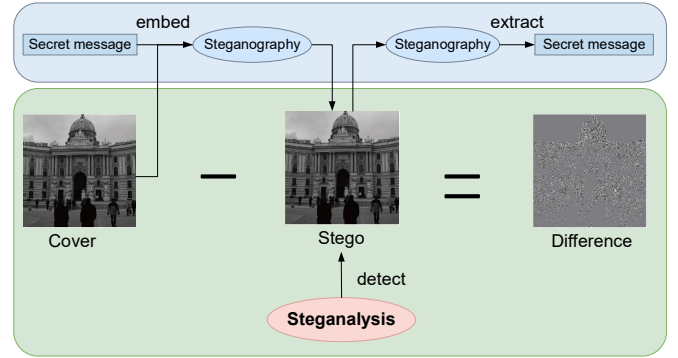

**Fig. 1.** Architecture of concealment communication. The stego generated by HUGO [2], payload is 0.4bpp, whose secret message is random binary data. Difference is the difference between the stego and the cover. White pixels mean that embedding modify pixel value is +1, and black pixels mean that embedding modify pixel value is -1.

## I. INTRODUCTION

STEGANALYSIS is used to embed secret information into media in invisible ways to achieve the goal of the privative conversation and digital copy right protection. Unfortunately, steganography can be used for illegal activities, such as spreading terrorism information, stealing secret, etc. For detecting and defensing this imperil, steganalysis is proposed to detect whether is secret embedding in digital media. Fig. 1 indicts the relationship between steganalysis and steganography.

Currently, the study of steganalysis still stays at binary classification stage to separate covers and stegos. Covers mean the images without secret message, stegos mean the images embedded with secret message. The difference image in Fig. 1 is the difference between the cover and the stego. Covers and stegos are similar in vision. Secret information embedding can be regarded as adding high-frequency and small amplitude noise on the cover [1]. Those noises call embedding signals. Because the secret message is embedded in high-frequency texture commonly, the embedding signals are covered by statistics feature of complex texture easily [3]. Therefore, it is challenging to extract signals. Currently, convolution kernel focuses on image semantics superiorly [4], hardly separates semantics edge and embedding signal. It causes embedding signals are sticky to be extracted and learned. Therefore, some special modifications are required for enhancing the embedding signal.

Present deep learning based steganalysis enhances embedding signal mainly by two ideas: reduce embedding signal loss and enhance embedding signal capture ability.

The approach of reducing embedding signal loss currently is to adjust network architecture, such as avoiding average pooling [5], introducing residual learning [6], which avoid the suppression of high-frequency signal. Average pool suppresses high-frequency signal including embedding signals. As the result, most of detectors discard average pooling in previse layers.

The enhancement of embedding signal capture ability mainly from adjustment of filters in the first layer. Traditional steganalysis methods regularly design multiplexed high-pass filters to generate residuals to extract high-frequency embedding signals regularly [7]. Deep learning models usually adjust filters referring the experience of traditional methods. According to reference forms, they can be categorize as fixed filters [8]–[10], learnable filters [11]–[13]. Fixed filters directly use the traditional high-pass filter weights. The learnable filters participate in backpropagation, which is widely applied because of its weight optimization advantage.

Almost all steganalysis methods are based on mainstream classification models. However, there is an enormous gap between steganalysis task and other CV task. Steganalysis is a low-SNR (signal noise ratio of stego signal to image) classification task [14]. But regular CV tasks are high-SNR (signal

Corresponding author: Songsen Yu.
H. Su, M. Han and S. Yu are with the School of Software, South China Normal University, Guangzhou, 510631, China (e-mail: suhai@m.scnu.edu.cn, hanmeiyin@m.scnu.edu.cn, yss8109@163.com).
J. Liang is with the College of Engineering & Computer Science, Syracuse University, Syracuse, NY 13244 USA (e-mail: juliang@syr.edu).

noise ratio of content to image) task. Each module of deep learning need to reevaluated to match the particularity of the steganalysis task. Current steganalysis focus on the filter and network architecture. We tried to explore modules that were ignored or could be improved. The paper concludes the disadvantages of modules from two aspects of reducing embedding signal loss and enhance embedding signal capture ability. They mainly include the following three points:

(1) *Activation function design:* Activation function impacts greatly in regular deep learning vision task. Most of steganalysis network selects ReLU as activation function in the later stage. In ordinary CV tasks, negative features are usually treated as unrelated noise such as background. ReLU avoids the interference of useless information with subsequent neurons. But, in steganalysis, embedding signal amplitude is significantly smaller than image content. While most steganalyzers use the ABS layer to set the residual to positive values, the possibility of presence of negative embedding signals in high-level features cannot be ruled out. ReLU discards all negative value, suppresses some of the embedding signals and affects the performance of steganalysis.

(2) *Filter design:* The success of SRM, traditional high-pass filters, is due to their special structure and diversity. Detectors with fixed filters try to combine multiple of unique high-pass filters and achieved accuracy improvement. While steganalyzers with learnable filters lacks the discussion of diversity. Traditional filters calculate the relationship between pixel and surrounding pixels in multi-dimension. However, current methods with learnable filters expand the range of filters into 5 × 5 during backpropagation. The diversity of residuals is difficult to reflect and the network hardly extract complex embedding signals.

(3) *Loss function design:* Loss function indicates the goal of deep learning task directly. Recently, most of detectors use cross entropy loss function directly, which is always used in classification task. However, because sematic edges and embedding signals are both high-frequency signal, extracted feature inevitably contains sematic edges. The high-frequency feature from the same class is various. Proved by ArcFace [15], cross entropy loss is not suitable for the classification task with large intra-class difference. Steganalysis trained by cross entropy loss is not able to separate embedding signal from Semantic features well.

In view of the problem above, we propose Steganalysis of Image with Adaptively Parametric Activation and our contributes are as follow:

(1) We propose Adaptively Parametric Activation Module (APAM), aiming at reducing embedding signal loss. APAM uses squeeze-and-excitation module to get threshold, which preserves the effective negative value.

(2) We design a new rule to restrain mathematical range of individual filter. Our filters preserve the diversity of residual and achieve the goal that network extracts more complex embedding signal.

(3) The paper introduces a loss function based on contrastive learning, which guides the network to pay attention to the discriminative embedding signals by maximizing the inter-class distance. It improves the ability to capture the embedding signals.

The remaining structure of this paper is consistent with four sections. Section 2 introduces briefly related work in the field of steganalysis. Section 3 describes the details of the method this paper proposed. In the Section 4, we analysis the result of experiments. Section 5 summarizes and looks forward to this article.

## II. RELATED WORK

The development of steganalysis can be divided into two main stages: traditional steganalysis and deep learning based steganalysis. Fig. 2 displays the three steps of steganalysis: computing residuals, feature representation, classification. In computing residuals, high-pass filters capture different types of dependencies between adjacent pixels. In feature representation, the network extracts statistical feature or high-level feature from residuals to gain embedding signal feature.

### A. Traditional steganalysis

Traditional steganalysis relies on handcraft feature to extract embedding signals. At first, traditional steganalyzers are required to design algorithm for specific steganography [16]–[19]. In order to strengthen versatility of steganalyzers, researchers study low-dimension features to capture embedding signals from stegos [20]. As the security of steganography raising, traditional steganalysis methods capture more hidden embedding signal by increasing feature dimension and diversity continuously. Those models are called as rich-model steganalysis. Currently, they perform well by designing diversiform linear and nonlinear filters, and its feature dimensions have reached more than tens of thousands of dimensions [21]–[23].

### B. Deep learning based steganalysis

The research of deep learning based steganalysis is still in its infancy, and the breadth and depth of research need to be further expanded. Given by powerful learning capabilities from deep learning, steganalysis can combine three steps to build an edge-to-edge steganalyzer. It gets a better performance comparing to rich-model steganalysis. In reality, the researchers find that it is useful to view the detector as three segments, although the architecture is edge-to-edge.

This paper introduces the CNN methods from the aspect of reducing embedding signal loss and enhance embedding signal respectively. To reduce embedding signal loss during forward propagation, current treatments are relatively single. They mostly only adjust the network structure. ZhangNet [24] replaces global averaging pooling by spatial pyramids to prevent feature loss. Both XuNet V3 [25] and YangNet [26] replace averaging pooling with convolutional layers whose stride is 2 to enhance the propagation of embedding signals. Besides, in order to prevent the signal loss, YangNet does not round pixel values to integers when decompressing JPEG images into a spatial domain.

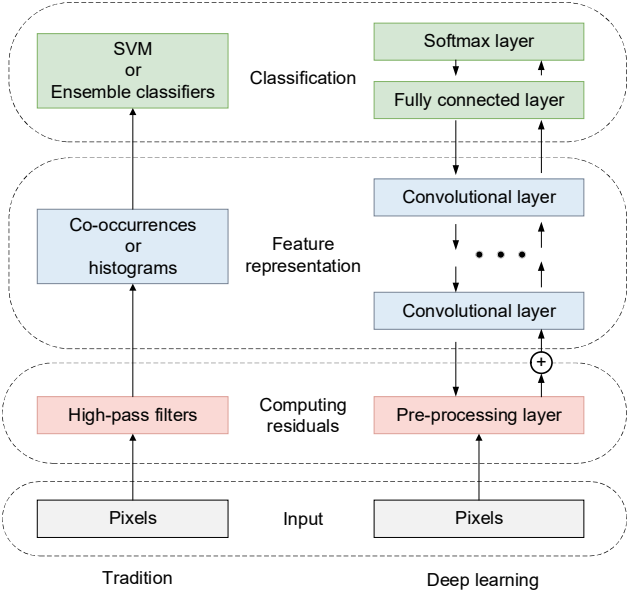

**Fig. 2.** The architecture of traditional steganalysis (left) and deep learning based steganalysis (right).

The study about improving the ability to extract embedding signals is relatively in-depth. We introduce it with following parts:

(1) *Filter design:* Deep learning based detectors not only adjust the training ways of filters to improve detect accuracy, but also enhance embedding signal by improving the diversity of filters. ReST-Net [27] compute residuals from sub-networks which utilizes three groups of high-pass filters. It indicates that multiple filters allow network to capture more embedding signals for both traditional and deep learning steganalysis.

(2) *Activation function Design:* Current discussion about activation function focusses on the first function after the filters. It simulates truncation in traditional steganalysis. ReST-Net selects TanH, which achieves the goal of truncation by limiting data distribution scale to [0,1]. YeNet [28] designs a novel activation function based on ReLU called truncated linear unit (TLU) by introducing a hyperparameter to restrict data scale manually.

(3) *Network design:* Researchers combine steganalysis knowledge and classification network to conduct network to learn and extract embedding signals. SiaStegNet [29] explores relationship of different sub-area in the same image by siamese network.

Based on the current study, the work of reducing embedding signal loss is not paid attention to, and the work of improving the ability to extract embedding signals still has unconsidered or unimproved modules. In this paper, we consider those insufficiencies in activation function, filters and loss function, to provide more idea to enhance embedding signal.

## III. PROPOSED METHOD

### A. Architecture

The keys of enhancing embedding signals are reducing embedding signal loss and enhancing embedding signals capture capability. We design an adaptively parametric activation module (APAM). It estimates amplitude of negative embedding signal to preserve the negative within threshold and reduce embedding signal loss. For improving embedding signal capture capability of the network, we propose a new restriction rule of filters and a loss function based on contrastive learning by increasing diversity of residuals and maximizing interclass distance and guiding network focuses on embedding signals.

According to the points above, the proposed architecture is shown in Fig. 3, following the mainstream architecture: computing residuals, feature representation, classification.

Computing residuals, whose responsibility is to suppresses image content and enhance embedding signals, generate residual image. In this step, a constraint rule is designed to maintain the diversity of filters by interfering the weight update of high-pass filters. Same as most steganalyzers, we insert TLU and ABS layer after filtering. TLU emulates traditional truncation to suppresses image content. ABS layer forces network to take into account the symbolic symmetry in the residual image. The paper utilizes same hyperparameter of TLU as YeNet.

Feature representation extract high-level feature from residuals. This step is consistent with 8 blocks, including 2 ReLU Blocks and 8 APAM Blocks. Each block goes through one convolution and activation sequentially. On former two layers, average pooling is avoided to reduce the embedding signal loss. At the meantime, we design APAM to preserve the signal within thresholds. Comparing to TLU, the proposed activation module is more suitable for input signal distribution in this stage.

In classification stage, in addition to the classification task, we execute the similarity task to calculate the similarity of examples in the feature space. It expends the interclass distance and guide the network to extract more embedding signals. The loss function based on contrastive learning, including cross entropy loss and pair-wise function, to calculate the error in two tasks. Cross entropy loss calculates the error of classification task and pair-wise function calculates the error of similarity task.

### B. Adaptive Parametric Activation Module

The embedding signals are easily lost during the forward propagation, which affects the detection accuracy. Most steganalysis models select ReLU as the activation function at the later layers, whose formula is shown in Eq. (1). YeNet shows that for other layers, the distribution of input signals tends to be similar to the one in CV tasks. Therefore, ReLU is more suitable in the later stages than TLU. The formula for TLU is shown in Eq. (2).

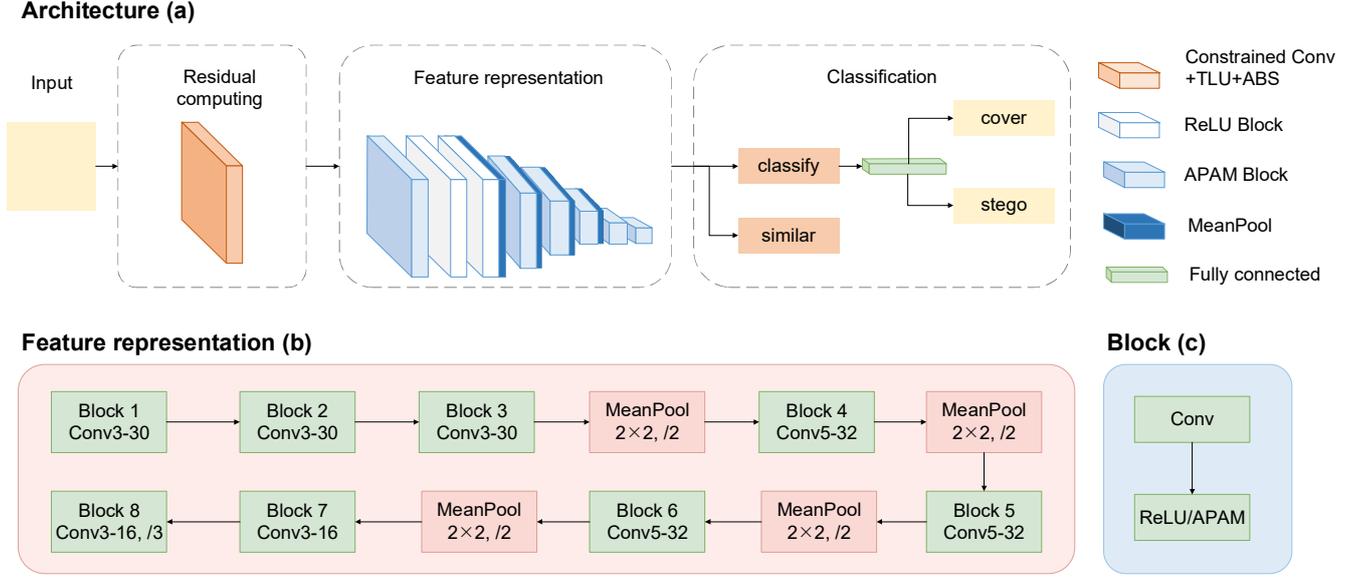

**Fig. 3.** Overview of our proposed network.

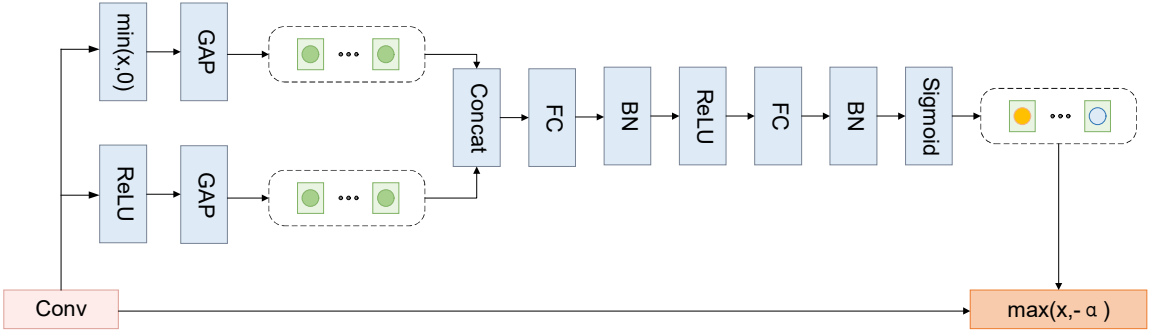

**Fig. 4.** Structure of APAM.

$$f_{ReLU}(x) = \begin{cases} x & if\ x > 0 \\ 0 & if\ x \leq 0 \end{cases} \quad (1)$$

$$f_{TLU}(x) = \begin{cases} T & if\ x > T \\ x & if\ -T \leq X \leq T \\ -T & if\ x < -T \end{cases} \quad (2)$$

However, ReLU potentially causes the loss of negative embedding signals. In regular CV tasks, negative signals generally are irrelative noises. ReLU avoids interference from useless information on subsequent neurons by transforming negative value to zero. However, in steganalysis, embedding signal is a kind of weak noise with small amplitude [28]. Even we apply ABS layer to change residuals feature to absolute value, it does means that there is no negative embedding signal in high-level feature. ReLU may loss the embedding signals when it suppresses the negative value. For stegos with low payload, a small loss of embedding signals greatly effects the detection accuracy.

In respect of the issues above, we design a new activation function based on ReLU to preserve the negative value and reduce the embedding signal loss. We only keep values greater than $-\alpha$, as shown in Eq. (3). Since the input signals of the other layers do not have the symbolic symmetry of residual features, they are more closely distributed to general CV tasks. So, differ from TLU, we do not suppress the positive value. It not only effectively reduces the loss of negative embedding signals, but also adapts to the distribution of input signals.

$$f_{AP}(x) = \begin{cases} x & if\ x > -\alpha \\ -\alpha & if\ x \leq -\alpha \end{cases} \quad (3)$$

We give $\alpha$ values of 0.01, 0.1, 0.5, and 0.8 according to experience. The improvement is not obvious, and some values even lead to a decline in accuracy. By analyzing experiment, the fixed hyperparameter is not advisable. Impacting from the image content of covers and stegos, the embedded traces of each stego are different. The small amplitude of embedding signals determines the sensitivity of detector to threshold. So, $\alpha$ needs to be flexible for the input image.

Enlightened by M. Zhao et al. [30] and H. Jie et al. [31],

TABLE I

INFLUENCE OF DIFFERENT BRANCHES ON THE ACCURACY.

|     | ReLU  | $min(x,0)$ | ReLU + $min(x,0)$ |
|-----|-------|------------|-------------------|
| Acc | 0.819 | 0.810      | **0.8401**        |

TABLE II

THE INFLUENCE OF DIFFERENT TRUNCATION DEGREES OF $x$ ON DETECTION ACCURACY

|     | $[-3, 0]$ | $[-2, 0]$ | $[-1, 0]$ | $[-\infty, 0]$ |
|-----|-----------|-----------|-----------|----------------|
| Acc | 0.8362    | 0.8331    | 0.8401    | **0.8411**     |

squeeze-and-excitation module is adopted. The module and our activation function constitute APAM. Fig. 4 shows structure of APAM. It is equivalent to calculating the amplitude of embedding signals. The feature map after convolution is input to ReLU branch and $min(x,0)$ branch respectively and propagated to GAP. Each branch gets 1D vectors representing the global information. Then, two 1D vectors are concatenated and go through the excitation module. Finally, the range of vectors are processed to $[0, 1]$ by sigmoid, which avoids the risk of assigning oversize value to activation. Furthermore, the range of $[0, 1]$ is reasonable, on the basis of the experiments with fixed parameters.

ReLU branch extracts the global positive information from feature maps, which contains numerous valuable information. $min(x, 0)$ branch extracts the global information of negative values from feature map. It provides availability information which is not contained in ReLU branch, such as negative embedding signals. The value of $\alpha$ is assigned by depending on both positive and negative globe information, which is relatively reliable.

The design motivation for APAM is as follows. First, since we need to activate the negative feature, the calculation of $\alpha$ requires the participation of negative information. Second, at first, we tried to directly extract the global features of feature map $x$ to calculate the threshold. But the effect is not the best. Therefore, negative information and positive information are processed separately. It further suggests that the two information are complementary. The specific experimental results are shown in TABLE I. Thirdly, in $min(x, 0)$ branch, we truncate negative values to different range, including $x \in [-3, 0]$, $x \in [-2, 0]$, $x \in [-1, 0]$. TABLE II shows that the change of accuracy is within 0.5%. It proves that the minimal negative value has little effect on the calculation, so we do not truncate the negative value. It is noted that the experimental models in this section is modified based on the final model.

Regarding the location of APAM, based on experiments, it is not idea to apply all APAM blocks. When Block 2 and Block 3 select ReLU, the model performs better. It may because the input signals in previous blocks contain lots of semantic statistical features, which gives rise to APAM incorrectly estimate the amplitude of embedding signals. The false estimation results in inputs of a lot of redundant information. Besides, it is essential to apply APAM in Block 1. Although the embedding signals are covered with complex sematic features, the suppression of negative value losses numerous available information. It effects the subsequent feature extraction. As the result, we apply ReLU on Block 2 and Block 3 and APAM on remaining blocks.

*C. Constrained Convolutional Layer*

Enhancing the capability of capturing embedding signals is one of the main factors to enhance embedding signals. The capture capability of recent steganalysis with learnable filters need to be improved. The decreased diversity of filter makes the hidden signals unable to be extracted. Based on the previous work, we formulate a constrain rule to fix the calculation scale of every filter, whose goal is to increase the diversity of residuals.

Filter, as an important part of steganalysis, extracts the dependency relationship between adjacent pixels. Diversified high-pass filters analyzes image from different angle, which enrich the extracted embedding signals [27].

Most of deep learning based steganalysis methods apply 30 high-pass filters with different calculated area and direction from SRM to initialize filters. These 30 high-pass filters explore the relationship between the central pixel and the surrounding different positions and quantity pixels, with a high degree of diversity. Fig. 5 shows that 30 filters correspond to 7 classes. Each class can be rotated to get filters in different directions, where class (a) includes 8 filters, class (b) includes 4 filters, class (c) includes 8 filters, class (d) includes 4 filters, class (f) includes 4 filters, class (e) includes one filter.

The diversity of filters is damaged due to training. All filters explore the dependency between center pixel and peripheral $5 \times 5$ pixels. The information in generated residual images become single. The diversity of residuals is the fundation of extracting rich embedding signals in later stage. Therefore, we hope to constrain the calculation range of the filter to increase the diversity of residuals.

Besides, larger range is not better. J. Fridrich et al. [32] find that pixel correlation is inversely proportional to the distance between pixels. The correlation on diagonal pixels decreases faster. It is essential to restrain the calculation scale properly. So, we setup different constraints for each filter to ensure the correlation is high-relative and diverse.

1) **Principle of filters**

Understanding the principle of filters is a prerequisite for designing reasonable constraint rules. In traditional steganalysis, SPAM [14], SRM, etc. are prediction residual features, which are generated by high-pass filters. They utilize a function $f(\cdot)$ to estimate the pixel in local windows. Then, the prediction residual $r$ is computed by subtraction of estimated value and real value showing as Eq. (4). This residuals calculation formulation, widely used in the field of steganalysis, builds local models of multi-pixel correlation to compute different residual features. The features are used to extract higher level feature.

$$r = f(I) - I \qquad (4)$$

where $I$ is the input image, $r$ is the residual image.

$$\begin{bmatrix} 0 & 0 & 0 & 0 & 0 \\ 0 & 0 & 1 & 0 & 0 \\ 0 & 0 & -1 & 0 & 0 \\ 0 & 0 & 0 & 0 & 0 \\ 0 & 0 & 0 & 0 & 0 \end{bmatrix} \quad \begin{bmatrix} 0 & 0 & 0 & 0 & 0 \\ 0 & 0 & 1 & 0 & 0 \\ 0 & 0 & -2 & 0 & 0 \\ 0 & 0 & 1 & 0 & 0 \\ 0 & 0 & 0 & 0 & 0 \end{bmatrix} \quad \begin{bmatrix} 0 & 0 & -1 & 0 & 0 \\ 0 & 0 & 3 & 0 & 0 \\ 0 & 0 & -3 & 0 & 0 \\ 0 & 0 & 1 & 0 & 0 \\ 0 & 0 & 0 & 0 & 0 \end{bmatrix} \quad \begin{bmatrix} 0 & 0 & 0 & 0 & 0 \\ 0 & -1 & 2 & -1 & 0 \\ 0 & 2 & -4 & 2 & 0 \\ 0 & -1 & 2 & -1 & 0 \\ 0 & 0 & 0 & 0 & 0 \end{bmatrix}$$
$$\quad\quad (a) \quad\quad\quad\quad\quad (b) \quad\quad\quad\quad\quad (c) \quad\quad\quad\quad\quad\quad (d)$$

$$\begin{bmatrix} 0 & 0 & 0 & 0 & 0 \\ 0 & -1 & 2 & -1 & 0 \\ 0 & 2 & -4 & 2 & 0 \\ 0 & 0 & 0 & 0 & 0 \\ 0 & 0 & 0 & 0 & 0 \end{bmatrix} \quad \begin{bmatrix} -1 & 2 & -2 & 2 & -1 \\ 2 & -6 & 8 & -6 & 2 \\ -2 & 8 & -12 & 8 & -2 \\ 0 & 0 & 0 & 0 & 0 \\ 0 & 0 & 0 & 0 & 0 \end{bmatrix} \quad \begin{bmatrix} -1 & 2 & -2 & 2 & -1 \\ 2 & -6 & 8 & -6 & 2 \\ -2 & 8 & -12 & 8 & -2 \\ 2 & -6 & 8 & -6 & 2 \\ -1 & 2 & -2 & 2 & -1 \end{bmatrix}$$
$$\quad\quad (e) \quad\quad\quad\quad\quad\quad (f) \quad\quad\quad\quad\quad\quad\quad\quad (g)$$

**Fig. 5.** Representative examples in 30 filters of SRM

$$\begin{bmatrix} 0 & 0 & 0 & 0 & 0 \\ 0 & 0 & 1 & 0 & 0 \\ 0 & 0 & -2 & 0 & 0 \\ 0 & 0 & 1 & 0 & 0 \\ 0 & 0 & 0 & 0 & 0 \end{bmatrix} \Longrightarrow \begin{bmatrix} 0 & 0 & 0 & 0 & 0 \\ 0 & 0 & 1 & 0 & 0 \\ 0 & 0 & 1 & 0 & 0 \\ 0 & 0 & 1 & 0 & 0 \\ 0 & 0 & 0 & 0 & 0 \end{bmatrix}$$

**Fig. 6.** The example of generating matrix $R$. The left is the original SRM filter, the right is generated matrix $R_k$.

Deep learning can simulate $f(\cdot)$ by convolution kernel to implement residual calculation. This method adaptively trains better filters to extract residual feature by backpropagation. The center weight is 0 during estimating. Assuming the center of $w_k$ is $(ctr, ctr)$, the new convolution kernel $\widetilde{w_k}$, which simulates $f(\cdot)$, is fitted to the following formula:

$$\widetilde{w_k} = \begin{cases} w_k(m,n) & if(m,n) \neq (ctr, ctr) \\ 0 & if(m,n) = (ctr, ctr) \end{cases} \quad (5)$$

Eq. (6) show the implementation of residual computation through convolution, which is similar to the traditional computation.

$$r = \widetilde{w_k} * I - I \quad (6)$$

where * means convolution operation.

2) **Constrained convolution**

We design a restrain rule to maintain the diversity of filters. The goal of the rule is to a ensure the pixels for prediction are same as the initial filters, which means that the calculation area is stable during training. Constrained convolution explores the relationship between center pixel and different neighboring pixel, which provides the diverse residual features for subsequent stages.

The reason why we keep filters identical to initial filters is that we find that the original non-zero region still maintain a large weight value after training. It proves the rationality of SRM design. Therefore, the calculation range is not redesign.

Matrix $R$, shown in Fig. 6, records the calculation area of each filter and its value is {0,1}. 1 means non-zero area, 0 means the original zero area.

After backpropagation, we utilized $R$ to limit the calculation area of filters, as shown in Eq. (7). Weights outside the area is 0.

$$w_k = w'_k * R_k \quad (7)$$

where $w'_k$ is the $k^{th}$ convolution kernel after backpropagation. $w_k$ is the $k^{th}$ filter after constraining. $R_k$ is the constraint matrix corresponding to the $k^{th}$ filter.

From B. Bayar et al. [4], the normalization of filters, which make the sum of filters is 1 excluding the center weight, improves the stability and convergence of filters, as Eq. (8) suggests. The center value is -1 after normalization. Thus, we normalize the filters.

$$\begin{cases} w_k(ctr, ctr) = -1 \\ \sum_{m,n \neq ctr} w_k(m,n) = 1 \end{cases} \quad (8)$$

Algorithm 1 shows the process of constrained convolution. We assume $w'_k$ is the filter after backpropagation. First, $w'_k$ is multiplied by $R$ and the outside weight value is set to 0. Then we set the center value of $w_k$ to 0 and normalize $w_k$ for estimating center value. Finally, the center value is set to -1, for fitting the formula of residual computation.

---
**Algorithm 1** Training Algorithm for Constrained Convolutional Layer
---
**Input:** The high-pass filter $k$, the weights of the $k$ high-pass filter $w'_k$
**Output:** The constrained weights $w_k$
1: Initialize $w'_k$ by SRM high-pass filters, $i = 1$
2: While $i \leq \text{max\_iter}$ do
3:  Do feedforward pass
4:  Update filter weights through stochastic gradient $w'_k$
5:   Descent and backpropagate errors
6:  Set $w_k = w'_k * R$
7:  Set $w_k(ctr, ctr) = 0$ for all K filters
8:  Normalize $w_k$ such that $\sum_{m,n \neq ctr} w_k(m,n) = 1$
9:  Set $w_k(ctr, ctr) = -1$ for all $k$ filter
10:  $i = i + 1$
11:  If training accuracy converges then
12:   Exit
13: End

---

Different pixel relationship is beneficial for extracting more potential signals. Comparing to current deep learning based steganalysis, our method limits the calculation area and forces filters calculate the correlation of pixels in fixed area to generated diverse residuals.

*D. Loss function based on contrastive learning*

In the previous section, the update of the weights is constrained to improve the diversity of filters. In this part, we optimize the loss function to guide the model to capture embedding signals from diverse semantic edges.

Most current deep learning steganalysis methods utilize cross entropy loss as the feedback information of back-propagation. We assume the label of covers is 0 and the label of stegos is 1.

The loss is implemented by a fully connection layer and a cross entropy loss function. The loss function is defined as Eq. (9).

$$L_{cls}(p, y) = \begin{cases} -\log(p) & if\ y == 1 \\ -\log(1-p) & if\ y == 0 \end{cases} \quad (9)$$

where $p \in [0,1]$ represents the output of the fully connection layer, which is predicted probability for the class, $y \in \{0,1\}$ represents ground truth label.

Cross entropy loss focuses on the output of fully connection layer and evaluates the distance between ground truth label and predicted label. The function does not explicitly optimize the feature to improve the similarity of intraclass and diversity of interclass. The generated feature is lack of discrimination, which causes poor performance when the appearance of intraclass is highly diversity [15]. In steganalysis task, it is hard to discriminate the sematic edges and high frequency embedding signals. When embedding signals are extracting, it is inevitable that the extraction is affected by sematic information. Stegos and covers contain the identical sematic information but the sematic information of intraclass is much different. Therefore, the intraclass of steganalysis is various in appearance and it is insufficient to train the model by cross entropy loss function.

In order to fit the steganalysis task, whose property is diverse of intraclass and similar of interclass, a pair-wise loss function [33] is introduced to descript the distance between high-level features vectors of covers and corresponding stegos. It can be viewed as a similarity task to compute similarity of examples. Pair-wise loss measures Euclidean distance between the vectors, showing as Eq. (10). $f_1$ and $f_2$ are the feature vector of covers and stegos respectively. $m$ is the manual hyperparameter. $y \in \{0,1\}$ represents the ground truth label, which 0 for covers, 1 for stegos. Through training, the distances of the intraclass are closer, and that of interclass are keep outside $m$.

$$L_{con}(f_1, f_2, y) = (1-y)\frac{1}{2}\|f_1 - f_2\|_2^2 + y\frac{1}{2}[\max(0, m - \|f_1 - f_2\|_2)]^2 \quad (10)$$

The embedding signal is the commonality of stegos, and it is also the key to distinguish stegos from covers. The quantity of embedding signals is inversely proportional to the distance of feature vectors of the intraclass. The model capture more embedding signals to shorten the distance under the penalty of loss, which is encouraged to distinguish between semantic edges and steganographic signals.

The final loss function based on contrastive learning is consist of cross entropy loss and pair-wise loss, combined with hyperparameter λ.

$$L = L_{cls}(p, y) + \lambda L_{con}(f_1, f_2, y) \quad (11)$$

## IV. EXPERIMENT RESULT

### A. The Steganographic Schemes and Datasets

We experimentalize on BOSSbase 1.01 [34]. The dataset contains 10,000 512 × 512 × 8 bit grayscale images with different texture features. Because of the high computation cost, the size of an image is resampled in half to improve the training speed [6], [13]. The research in [28] proved different down-sampling methods have influence on the detection performance. Therefore, we uniformly use imresize() with default parameter in Matlab.

Same as most deep learning based steganalysis, we do not detect deep learning steganography because it is still at an embryonic stage [29]. Our datasets are made by WOW [35] and S-UNIWARD [36] methods (codes from http://dde.binghamton.edu/download/), whose payloads respectively are 0.2 bpp, 0.4 bpp. Eq. (12) shows the formula of payload. $n_{secret}$ represents the number of bits of the secret information and $n_{secret}$ represents the number of bits of the cover. The datasets are divided into three parts, 40% cover-stego pairs for training, 10% pairs for validation and the rest for testing.

$$payload = \frac{n_{secret}}{n_{cover}} \quad (12)$$

### B. Experiment sets

We utilize AdaDelta [37] to train our network with the mini-batch size of 32, containing 16 cover-stego pairs. The momentum and the weight decay of networks are set to 0.95 and $5 \times 10^{-4}$ respectively. Parameter "delta" of AdaDelta is $1 \times 10^{-8}$. For filters in the computing residuals stage, the weights are initialized with 30 filters in SRM, which participate in the weight update during training. In addition, StepLR is adopted in the paper, and the learning rate decreased by 20% every 50 epochs.

It is worth noting that in the experiments, no identical cover exists among the training set, verification set and test set. Same as SRNet, considering the difficulty of convergence in low payload datasets, we adopted transfer learning. The network is trained for 0.4bpp payload first and then trained in the order of "0.2←0.3←0.4".

We set up a series of experiments, which are summarized as follows:

(1) *Hyperparameter experiment of loss function:* The influence of different values of hyperparameters in our loss function is discussed.

(2) *The necessity of adaptive parameters in APAM:* By comparing the performances of fixed parameters and adaptive parameters, we discuss the necessity of adopting adaptive parameters.

(3) *Experiments of constraint rules:* In order to study the influence of diversity on the model and prove the advantage of the proposed constraint rules, three different rules are set to constrain the filters.

(4) *Ablation experiments:* Observe the performance impact of each component on the model.

TABLE III

DETECTION ACCURACY OF THE NETWORK WITH DIFFERENT M AGAINST S-UNIWARD AT 0.4BPP

|  | m = 2 | m = 2.5 | m = 3 | m = 3.5 | m = 4 |
|---|---|---|---|---|---|
| Acc | 0.822 | 0.815 | **0.841** | 0.837 | 0.826 |

TABLE IV

DETECTION ACCURACY OF THE NETWORK WITH DIFFERENT $\lambda$ AGAINST S-UNIWARD AT 0.4 BPP

|  | $\lambda = 0$ | $\lambda = 0.01$ | $\lambda = 0.05$ | $\lambda = 0.1$ | $\lambda = 0.5$ |
|---|---|---|---|---|---|
| Acc | 0.808 | 0.830 | **0.841** | 0.824 | 0.828 |

TABLE V

AT 0.4 BPP, THE ACCURACY OF THE NETWORK WITH FIXED PARAMETER AND ADAPTIVE PARAMETER AGAINST S-UNIWARD.

|  | $\alpha = 0.01$ | $\alpha = 0.1$ | $\alpha = 0.5$ | $\alpha = 0.8$ | Ours |
|---|---|---|---|---|---|
| Acc | 0.729 | 0.744 | 0.745 | 0.729 | **0.841** |

TABLE VI

IN THE FIRST 100 IMAGES, THE MEAN AND STANDARD DEVIATION OF THRESHOLD VALUES OF EACH APAM BLOCK, AND THE NUMBERS IS THE POSITION OF BLOCKS IN THE FEATURE REPRESENTATION STAGE IN FIG. 3.

|  | APAM1 | APAM4 | APAM5 |
|---|---|---|---|
| Mean | 0.5120 | 0.4972 | 0.4933 |
| Std | 0.0441 | 0.0209 | 0.0320 |
|  | APAM6 | APAM7 | APAM8 |
| Mean | 0.4895 | 0.4999 | 0.4982 |
| Std | 0.0524 | 0.0004 | 0.0957 |

(5) *Applicability of APAM:* We explore situations in which APAM is effective. Meanwhile, we try to add APAM to the mainstream steganalyzers.

(6) *Compare with existing methods:* Explore the superiority of the proposed method over existing methods. Because pf various experimental settings, we only select methods with identical experimental settings for comparison.

(7) *The T-SNE visualization:* We visualized the convolutional layer at the end of the network to check the distribution of extracted features.

*C. Hyperparameter experiment of loss function*

There are two hyperparameters in our loss function: the distance $m$ between inter-classes in contrastive learning and the ratio $\lambda$ of the cross-entropy loss and contrastive loss.

In TABLE III, five values m are set in the experiment. According to the results, the larger distance is not the better. This is because when the network tries to optimize the distance of feature vector pairs to an impossible value, examples that are difficult to detect will not get attention. When the distance is a reasonable value, after a certain number of epochs, we can no

TABLE VII

ACCURACY OF THE NETWORK TRAINED BY DIFFERENT CONSTRAINT RULES AGAINST S-UNIWARD AT 0.4 BPP

|  | Our constraint | Direction constraint | Without constraint |
|---|---|---|---|
| Acc | **0.841** | 0.819 | 0.791 |

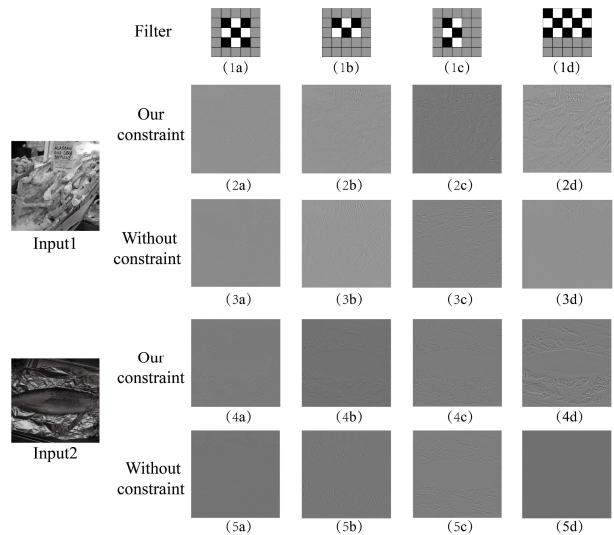

**Fig. 7.** Visualize residual images generated by four filters. "Filter" is the illustration of the weight of the initialized high-pass filters. Black represents weight less than 0, white represents weight greater than 0, and gray represents weight 0. The filters initialized with the same high-pass filter for each column. (2a) - (2d) and (4a) - (4d) are generated by our filters. Residual images of (3a) -(3D) and (5A) - (5D) are generated by unconstrained filters.

longer pay attention to the examples that can be identified correctly. According to the experimental results, 3 is the ideal distance.

TABLE IV shows the effect of $\lambda$ on the model. When $\lambda = 0$, the loss function is the cross-entropy function, and the effect is the worst. When $\lambda = 0.05$, the detection accuracy is the highest. This may be because contrastive loss and cross entropy loss can be used as complementary loss functions. Contrastive loss focuses on the feature representation of the network and cross-entropy loss focuses on the mapping of features to discrete labels. So, that each is irreplaceable.

*D. The necessity of adaptive parameters in APAM*

This experiment compares the effects of fixed and adaptive parameters in APAM. The parameter $\alpha$ determines the range in which negative values are retained. We manually set 4 kinds of hyperparameters as threshold α. According to the results of TABLE V, the improvement of fixed $\alpha$ is weak and that of adaptive $\alpha$ is obvious.

TABLE VIII

THE ACCURACY OF NETWORK TRAINED BY DIFFERENT COMPONENTS AGAINST S-UNIWARD AT 0.4BPP.

|     | Origin | APAM | Constraint Conv |
|-----|--------|------|-----------------|
| Acc | 0.561  | 0.562 | 0.673          |
|     | Contrastive Learning | APAM + Constraint Conv | APAM + Contrastive Learning |
| Acc | 0.700  | 0.756 | 0.700          |
|     | Constraint Conv + Contrastive Learning | | Ours |
| Acc | 0.702  | | **0.841**     |

TABLE IX

THE MEAN PROPORTION OF WHITE DOTS IN THE POSITIVE AND NEGATIVE AREAS OF 100 PAIRS OF IMAGES FROM THE ACTIVATION FUNCTION INPUT OF BLOCK 1.

| | Group A | |
|---|---|---|
| | Origin | Contrastive learning |
| Positive | 41.50% | 36.12% |
| Negative | 17.24% | 19.93% |
| | Group B | |
| | Constraint conv | Constraint conv |
| Positive | 34.81% | 34.81% |
| Negative | 30.36% | 30.36% |

We take a closer look at the thresholds calculated by the APAM. According to the results in TABLE VI, the average thresholds of all APAM blocks are close to 0.5. Most thresholds are in the range of 0.45 to 0.55 based on standard deviation values. But the improvement of using fixed parameter 0.5 is very unsatisfied. It indicates that a small threshold fluctuation has an impact on model performance, which may be caused by the low amplitude characteristics of embedding signals.

Based on the above findings, adaptive parameters are more suitable for characteristics of steganalysis. Therefore, it is necessary to utilize adaptive parameters.

*E. Experiments of constraint rules*

The paper mainly discusses the validity of the proposed constraint rule in the filters. This experiment mainly compares three constraint rules :(1) "Our constraint" only preserves the non-zero region of the initialized high-pass filter without saving symbols; (2) "Direction constraint" preserves not only non-zero region but also symbols; (3) "Without constraint" means not to impose constraints.

The experimental results in this part are based on the accuracy of steganalyzer trained by S-UNIWARD with 0.4bpp. According to TABLE VII, "Our constraint" is the best. The accuracies of "Our constraint" and "Direction constraint" are better than that of "Without constraint". It proves the gain of multiple residuals on steganalysis performance.

In order to further demonstrate the advantages of constraint rules in the constrained convolution layer, we visualize the residual image after filtering. The effect of the filters with large kernel is obvious, so we only show part of the filters. The "Filter" in Fig. 8 is the high-pass filter for initialization of this column. There are two findings we can see from the visualization. First, the residual image generated by our filters have finer features. For example, Fig. 7 (2a) contains more texture than Fig. 7 (3a). Second, the proposed filters are more directional. For example, Fig. 7 (2b) extracts more vertical edges of the input image than Fig. 7 (3b). This is because our filters better preserve the horizontal direction of the original high-pass filter.

*F. Ablation experiments*

In this part, we explore the influence of each component on detection accuracy through experiments. The dataset in this experiment is generated by S-UNIWARD with 0.4bpp. Observing the TABLE XIV, we find both "Constraint Conv" and "Contrastive Learning" can improve the detection accuracy. It indicates that improving the ability of extracting embedding signals can bring gains to steganalysis.

In addition, compared with "Contrastive Learning", the accuracy of "APAM + Contrastive Learning" also has not improved. However, for "APAM + Constraint Conv", the increase in accuracy is significant. It shows that APAM is not entirely suitable for all networks. The applicability of APAM will be discussed in later experiment.

*G. Applicability of APAM*

In ablation experiments, we find that APAM does not applicable to all networks. We refer to the networks that are not suitable for APAM as group A, including "Origin" (without using the proposed methods), and "Contractive Learning." The network suitable for APAM is called Group B, including "Constraint Conv", "Constraint Conv +Contrastive Learning". We try to explore the applicability of APAM.

1) **Selection of the indicator**

Since the design of APAM is based on the consideration of effective negative value, we visualized the feature map of activation function input to observe the change of negative value area under different network. In order to explore whether negative values are valid information, the difference between feature maps of cover and its corresponding stego is calculated. When the pixel difference value is greater than a threshold, we can consider that it may be the effective information to distinguish cover and stego. It can be considered the embedding signal. This is because after filtering, the feature map mainly includes embedding signals, semantic edges and redundant noises (such as the prediction error of high-pass filters). Cover and stego have the same semantic edge, so the difference between the se mantic edge is approximately 0. When the threshold is a reasonable value, the pixel whose difference is greater

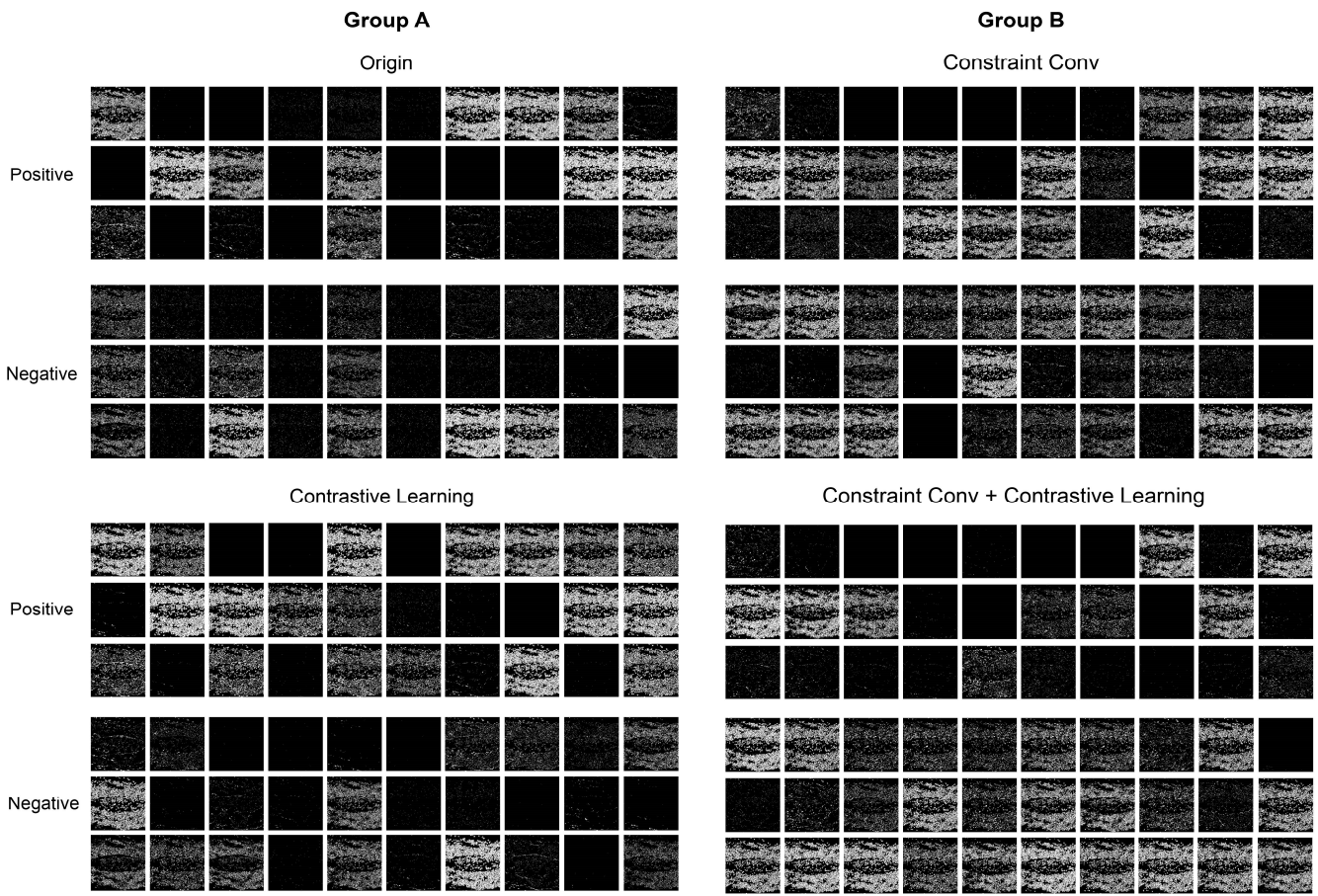

**Fig. 8.** In the activation function input of Block 1, the difference between feature maps of cover and its corresponding stego is visualized. The input image is the Input2 in Fig. 7. The white dots indicate that the pixel difference here is greater than the average difference of the image. Positive and negative respectively indicate the positive and negative area of the cover.

TABLE X

THE MEAN PROPORTION OF WHITE DOTS IN THE POSITIVE AND NEGATIVE AREAS OF 100 PAIRS OF IMAGES FROM THE INPUT OF FIRST ReLU.

|  | YeNet | XuNet V1 |
|---|---|---|
| Positive | 35.55% | 33.77% |
| Negative | 29.80% | 38.13% |

TABLE XI

AT 0.4 BPP, THE ACCURACY OF YENET WITH APAM AND XUNET WITH APAM AGAINST S-UNIWARD. "+N%" MEANS N% HIGHER THAN THE ACCURACY WITHOUT APAM.

|  | YeNet+APAM | XuNet V1+APAM |
|---|---|---|
| Acc | 0.790(+6%) | 0.783(+3.1%) |

TABLE XII

ACCURACY OF SIX STEGANALYZERS AGAINST WOW AND S-UNIWARD AT 0.2 BPP AND 0.4 BPP.

|  | WOW | | S-UNIWARD | |
|---|---|---|---|---|
|  | 0.2 bpp | 0.4 bpp | 0.2 bpp | 0.4 bpp |
| SRM+EC (2012) | 0.668 | 0.759 | 0.654 | 0.766 |
| XuNet V1 (2016) | 0.676 | 0.785 | 0.635 | 0.752 |
| YeNet (2017) | 0.714 | 0.812 | 0.631 | 0.773 |
| YedroudjNet (2018) | 0.723 | 0.831 | 0.630 | 0.781 |
| SRNet (2019) | 0.754 | 0.869 | 0.674 | 0.816 |
| ZhuNet (2020) | 0.767 | **0.882** | 0.715 | **0.847** |
| DFSE-Net (2021) | 0.753 | 0.851 | 0.659 | 0.785 |
| Ours | **0.802** | 0.874 | **0.770** | 0.841 |

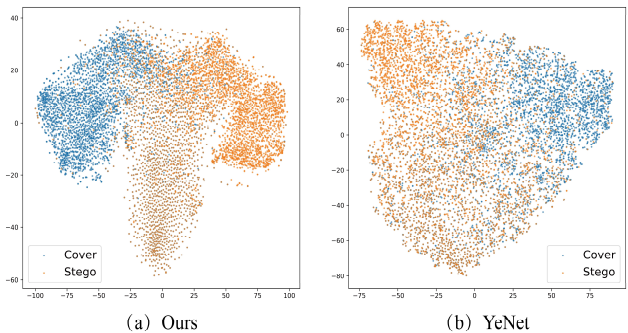

**Fig. 9.** T-SNE visualization of high-level features.

than the threshold is only embedding signals or redundant noises. It is difficult to set an appropriate threshold artificially because each input has a slightly different range of value. In this paper, we choose the average of the global difference as the threshold. This is based on the following two points: (1) Due to the low payload of traditional steganography, only a few pixels are modified. Most of the difference should be equal to or close to 0. (2) After the truncation operation of TLU, the abnormally large values caused by the prediction error have been sup pressed and do not affect the mean. Therefore, when the pixel difference is greater than the average, the difference can be considered to be large. At this point, the area may contain valid information.

We distinguish between positive and negative areas according to the feature map of cover. Because the cover is an unmodified image, its pixel value is considered a reference value. Therefore, the feature map of cover is the benchmark. We only explore the activation function closest to the computing residual stage. At this time, the feature map is similar to the input image and residual image, which is convenient for observation.

2) **The situation of the proposed network**

Fig. 8 is the result of our visualization of Group A and Group B, and the input image is Input2 of Fig. 7. The white dots of positive area represent the pixels whose difference is greater than the threshold in the Positive area of cover. The white dots in the negative region have a similar meaning. The more white dots, the more likely the area is to contain valid information. Certainly, the existence of redundant noise cannot be ruled out. According to the results, compared with the input image, white dots are mainly distributed in the texture region. The number of white dots in the negative area of group A is obviously lower than that in the positive area. Moreover, the number of white dots in the negative area of group B is obviously close to or greater than that in the positive area. Therefore, the negative region of the model in group B is more likely to have valid information than that in group A.

We calculated the proportion of white dots in the positive and negative areas of 100 pairs of images. The average proportion is the number of white dots divided by the total number of the pixel. It can be seen from TABLE VIII that the white dots of negative regions in group A is generally more than that in group B. In addition, they are closer to that of positive regions. Combined with the results of ablation experiment, APAM added in group B has a better effect, while added in group A has no obvious improvement. The addition of APAM enables the negative valid information in group B to be processed. The probability of the group A having effective negative values is low, so APAM does not enhance the performance of steganalyzer.

This seems to explain why APAM is sometimes useful and sometimes not. In order to avoid the influence of the uniqueness of our models, we further experiment on some mainstream steganalyzers.

3) **The performance in the mainstream detectors**

We extract the input of the first ReLU in YeNet and XuNet. The mean proportion of 100 pairs of images is calculated. According to the results of TABLE X, there are many pixels with large differences in the negative areas of YeNet and XuNet, which may contain effective negative values. We reimplement and train YeNet and XuNet by replacing ReLu with APAM. The improvement of accuracy is shown as TABLE XI. APAM improves the detection capability of YeNet and XuNet.

In the above experiments, the paper explores the applicability of APAM. We can estimate whether there are valid negative values in the steganalyzer according to the method in the paper. Then we can try to use APAM to process negative values to improve the detection ability.

*H Compared with existing methods*

In order to verify the superiority of our method, we compare SRM+EC [32], [38], XuNet [39], YeNet [28], YedroudjNet [9], SRNet [8], ZhuNet [24] and DFSE-NET [40]. The results of SRM+EC, XuNet, YeNet, YedroudjNet and SRNet are taken from ZhuNet. According to

TABLE XII, the proposed method performs best for most of steganalyzers under various payloads. It proves the effectiveness of the proposed strategy in reducing embedding signal loss and improving embedding signal extraction.

Although in the case of 0.4 bpp, the accuracy of our method is only close to that of ZhuNet, our steganalyzer performs significant better in detecting the stego with low payload. It may mainly due to the strategy of reducing negative embedding signal loss. The stego with low payload contains weaker embedding signals. However, in the process of forward propagation, we try to retain the extracted embedding signals, which provides a more reliable information for the final classification.

*I. The T-SNE visualization*

T-SNE [41] is used to visualize the high-level features of the output of the last convolutional layer to check the distribution of examples in the feature space. The visualized objects are stegos generated by S-UNIWARD with 0.4bpp and its corresponding covers.

According to the visualization results, we can estimate the final extraction of embedding signals. When there are obvious class clusters and the boundary between class clusters, it is proved that the high-level feature is distinctive and the feature contains more embedding signals. Depending on Fig. 9, compared with YeNet, our method has apparent cover and stego class clusters. There is a clear boundary between the centers of the two class clusters. The improvement of visualization results is attributed to our strategies, especially the penalty of minimizing the intra-class distance and maximizing the inter-class distance by the loss function based on contrastive learning.

## V. CONCLUSION

Steganalysis of Image with Adaptively Parametric Activation are proposed in this paper. From two aspects of reducing embedding signal loss and improving embedding signal extraction ability, we explore the design defects of activation function, filter and loss function and propose a novel strengthen embedding signal scheme. We design APAM to estimate the amplitude of negative embedding signals and reduce the loss of effective negative values. We also utilize two methods to improve the ability of extracting embedding signals: (1) The constraint rule is designed to restrict the residual calculation area of filters to maintain the diversity of residuals. (2) The loss function based on contrastive learning is introduced to distinguish embedding signals from semantic edges through the punishment of maximizing inter-class distance. The compare experiment proves that our strategy can improve the detection ability, especially the detection ability of image with low payload. In addition, the ablation experiment reflects the gain of each proposed scheme. At the same time, we explore the applicability of APAM in different models to test the universality of APAM.

The paper explores the validity of negative values in steganalysis and improves detection ability by retaining some of the negative. In the future, we will further explore the characteristics of negative values to better serve our detector.

## REFERENCES

[1] S. Wu, S. H. Zhong, and Y. Liu, "A Novel Convolutional Neural Network for Image Steganalysis with Shared Normalization," *IEEE Transactions on Multimedia*, 2017.

[2] T. Pevný, T. Filler, and P. Bas, "Using High-Dimensional Image Models to Perform Highly Undetectable Steganography," *Lecture Notes in Computer Science*, vol. 6387, pp. 161–177, 2010.

[3] Y. Qian, J. Dong, W. Wang, and T. Tan, "Feature learning for steganalysis using convolutional neural networks," *Multimed Tools Appl*, vol. 77, no. 15, pp. 19633–19657, Aug. 2018, doi: 10.1007/s11042-017-5326-1.

[4] B. Bayar and M. C. Stamm, "Constrained Convolutional Neural Networks: A New Approach Towards General Purpose Image Manipulation Detection," *IEEE Trans.Inform.Forensic Secur.*, vol. 13, no. 11, pp. 2691–2706, Nov. 2018, doi: 10.1109/TIFS.2018.2825953.

[5] M. Boroumand, M. Chen, and J. Fridrich, "Deep Residual Network for Steganalysis of Digital Images," *IEEE Trans.Inform.Forensic Secur.*, vol. 14, no. 5, pp. 1181–1193, May 2019, doi: 10.1109/TIFS.2018.2871749.

[6] S. Wu, "Deep residual learning for image steganalysis," *Multimed Tools Appl*, p. 17, 2018.

[7] J. Fridrich, J. Kodovský, V. Holub, and M. Goljan, "Steganalysis of Content-Adaptive Steganography in Spatial Domain," in *Proceedings of the 13th International Conference on Information Hiding*, Berlin, Heidelberg, 2011, pp. 102–117.

[8] Y. Qian, J. Dong, W. Wang, and T. Tan, "Deep learning for steganalysis via convolutional neural networks," San Francisco, California, United States, Mar. 2015, p. 94090J. doi: 10.1117/12.2083479.

[9] M. Yedroudj, F. Comby, and M. Chaumont, "Yedroudj-Net: An Efficient CNN for Spatial Steganalysis," in *2018 IEEE International Conference on Acoustics, Speech and Signal Processing (ICASSP)*, Calgary, AB, Apr. 2018, pp. 2092–2096. doi: 10.1109/ICASSP.2018.8461438.

[10] J. Zeng, S. Tan, B. Li, and J. Huang, "Large-Scale JPEG Image Steganalysis Using Hybrid Deep-Learning Framework," *IEEE Trans.Inform.Forensic Secur.*, vol. 13, no. 5, pp. 1200–1214, May 2018, doi: 10.1109/TIFS.2017.2779446.

[11] S. Tan and B. Li, "Stacked convolutional auto-encoders for steganalysis of digital images," in *Signal and Information Processing Association Annual Summit and Conference (APSIPA), 2014 Asia-Pacific*, Chiang Mai, Thailand, Dec. 2014, pp. 1–4. doi: 10.1109/APSIPA.2014.7041565.

[12] Y. Yousfi and J. Fridrich, "An Intriguing Struggle of CNNs in JPEG Steganalysis and the OneHot Solution," *IEEE Signal Process. Lett.*, vol. 27, pp. 830–834, 2020, doi: 10.1109/LSP.2020.2993959.

[13] M. Chen, M. Boroumand, and J. Fridrich, "Reference Channels for Steganalysis of Images with Convolutional Neural Networks," in *Proceedings of the ACM Workshop on Information Hiding and Multimedia Security*, Paris France, Jul. 2019, pp. 188–197. doi: 10.1145/3335203.3335733.

[14] T. Pevny, P. Bas, and J. Fridrich, "Steganalysis by Subtractive Pixel Adjacency Matrix," *IEEE Transactions on Information Forensics & Security*, vol. 5, no. 2, pp. 215–224, 2010.

[15] J. Deng, J. Guo, N. Xue, and S. Zafeiriou, "ArcFace: Additive Angular Margin Loss for Deep Face Recognition," *arXiv:1801.07698 [cs]*, Feb. 2019, Accessed: Nov. 02, 2021. [Online]. Available: http://arxiv.org/abs/1801.07698

[16] W. Bender, D. Gruhl, N. Morimoto, and A. Lu, "Techniques for data hiding," *Ibm Systems Journal*, vol. 35, no. 3.4, pp. 313–336, 1996.

[17] F. A. P. Petitcolas, R. J. Anderson, and M. G. Kuhn, "Information hiding-a survey," *Proceedings of the IEEE*, vol. 87, no. 7, pp. 1062–1078, 1999.

[18] X. Zhang, S. Wang, and K. Zhang, "Steganography with Least Histogram Abnormality," 2003.

[19] X. Luo, Z. Hu, C. Yang, and S. Gao, "A secure LSB steganography system defeating sample pair analysis based on chaos system and dynamic compensation," 2006.

[20] Y. Q. Shi, C. Chen, and W. Chen, "A Markov process based approach to effective attacking JPEG steganography," in *International Workshop on Information Hiding*, 2006, pp. 249–264.

[21] V. Holub and J. Fridrich, "Random Projections of Residuals for Digital Image Steganalysis," *IEEE Transactions on Information Forensics and Security*, vol. 8, no. 12, p. 1996, 2013.

[22] X. Song, F. Liu, C. Yang, X. Luo, and Y. Zhang, "Steganalysis of Adaptive JPEG Steganography Using 2D Gabor Filters," in *Proceedings of the 3rd ACM Workshop on Information Hiding and Multimedia Security*, New York, NY, USA, 2015, pp. 15–23. doi: 10.1145/2756601.2756608.

[23] J. Yu, F. Li, H. Cheng, and X. Zhang, "Spatial Steganalysis Using Contrast of Residuals," *IEEE Signal Processing Letters*, pp. 989–992, 2016.

[24] R. Zhang, F. Zhu, J. Liu, and G. Liu, "Depth-Wise Separable Convolutions and Multi-Level Pooling for an Efficient Spatial CNN-Based Steganalysis," *IEEE Trans.Inform.Forensic Secur.*, vol. 15, pp. 1138–1150, 2020, doi: 10.1109/TIFS.2019.2936913.

[25] G. Xu, "Deep convolutional neural network to detect J-UNIWARD," in *Proceedings of the 5th ACM Workshop on Information Hiding and Multimedia Security*, 2017, pp. 67–73.

[26] J. Yang, X. Kang, E. K. Wong, and Y.-Q. Shi, "Deep learning with feature reuse for JPEG image steganalysis," in *2018 Asia-Pacific Signal and Information Processing Association Annual Summit and Conference (APSIPA ASC)*, 2018, pp. 533–538.

[27] B. Li, W. Wei, A. Ferreira, and S. Tan, "ReST-Net: Diverse Activation Modules and Parallel Subnets-Based CNN for Spatial Image Steganalysis," *IEEE Signal Process. Lett.*, vol. 25, no. 5, pp. 650–654, May 2018, doi: 10.1109/LSP.2018.2816569.

[28] J. Ye, J. Ni, and Y. Yi, "Deep Learning Hierarchical Representations for Image Steganalysis," *IEEE Trans.Inform.Forensic Secur.*, vol. 12, no. 11, pp. 2545–2557, Nov. 2017, doi: 10.1109/TIFS.2017.2710946.

[29] W. You, H. Zhang, and X. Zhao, "A Siamese CNN for Image Steganalysis," *IEEE Trans.Inform.Forensic Secur.*, vol. 16, pp. 291–306, 2021, doi: 10.1109/TIFS.2020.3013204.

- [30] M. Zhao, S. Zhong, X. Fu, B. Tang, S. Dong, and M. Pecht, "Deep Residual Networks With Adaptively Parametric Rectifier Linear Units for Fault Diagnosis," *IEEE Trans. Ind. Electron.*, vol. 68, no. 3, pp. 2587–2597, Mar. 2021, doi: 10.1109/TIE.2020.2972458.
- [31] H. Jie, S. Li, S. Gang, and S. Albanie, "Squeeze-and-Excitation Networks," *IEEE Transactions on Pattern Analysis and Machine Intelligence*, vol. PP, no. 99, 2017.
- [32] J. Fridrich and J. Kodovsky, "Rich Models for Steganalysis of Digital Images," *IEEE Transactions on Information Forensics and Security*, vol. 7, no. 3, pp. 868–882, 2012, doi: 10.1109/TIFS.2012.2190402.
- [33] R. Hadsell, S. Chopra, and Y. Lecun, "Dimensionality Reduction by Learning an Invariant Mapping," 2006.
- [34] P. Bas, T. Filler, and T. Pevn, "'Break Our Steganographic System': The Ins and Outs of Organizing BOSS," 2011.
- [35] V. Holub and J. Fridrich, "Designing steganographic distortion using directional filters," in *2012 IEEE International Workshop on Information Forensics and Security (WIFS)*, 2012, pp. 234–239. doi: 10.1109/WIFS.2012.6412655.
- [36] V. Holub, J. Fridrich, and T. Denemark, "Universal distortion function for steganography in an arbitrary domain," *Eurasip Journal on Information Security*, vol. 2014, no. 1, p. 1, 2014.
- [37] M. D. Zeiler, "ADADELTA: An adaptive learning rate method," *Computer ence*, 2012.
- [38] J. Kodovsky, J. Fridrich, and V. Holub, "Ensemble Classifiers for Steganalysis of Digital Media," *IEEE Transactions on Information Forensics & Security*, vol. 7, no. 2, pp. 432–444, 2012.
- [39] G. Xu, H.-Z. Wu, and Y.-Q. Shi, "Structural Design of Convolutional Neural Networks for Steganalysis," *IEEE Signal Process. Lett.*, vol. 23, no. 5, pp. 708–712, May 2016, doi: 10.1109/LSP.2016.2548421.
- [40] F. Liu, X. Zhou, X. Yan, Y. Lu, and S. Wang, "Image Steganalysis via Diverse Filters and Squeeze-and-Excitation Convolutional Neural Network," 2021.
- [41] L. Van der Maaten and G. Hinton, "Visualizing data using t-SNE.," *Journal of machine learning research*, vol. 9, no. 11, 2008.

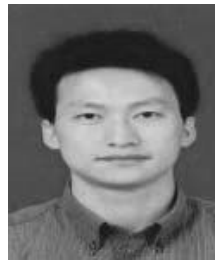

**Songsen Yu** received the Ph.D. degree in Guangdong University of Technology. Since 2009, he has been with the School of Software, South China Normal University, South China Normal University, where he is currently a professor. His research interests include Internet of Things, big data and artificial intelligence.

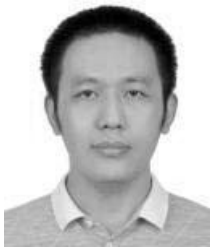

**Hai Su** received the B.S. and Ph.D. degrees in Graphic Communication from the School of Printing and Packing, Wuhan University, Wuhan, China, in 2010 and 2015, respectively. He is currently a Lecturer with the School of Software, South China Normal University, Guangzhou, China. His research interests mainly include image retrieval and image steganography.

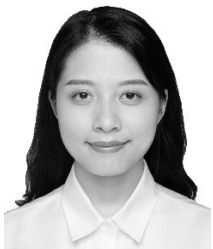

**Meiyin Han** received the B.S. degree from South China Normal University, Guangzhou, China, in 2016 and 2020. She is currently pursuing the M.S. degree with the School of Software in South China Normal University. Her research interests include image retrieval and steganalysis.

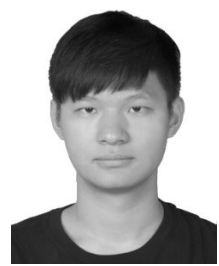

**Junle Liang** received the B.S. degree from South China Normal University in 2016. He is currently pursuing the M.S. degree with the Syracuse University of College of Engineering & Computer Science. His research interests include image retrieval and machine learning.